\documentclass[twocolumn]{aastex63}

\shorttitle{The Dynamical Viability of an Extended Jupiter Ring
  System}
\shortauthors{Stephen R. Kane \& Zhexing Li}

\begin{document}

\title{The Dynamical Viability of an Extended Jupiter Ring System}

\author[0000-0002-7084-0529]{Stephen R. Kane}
\affiliation{Department of Earth and Planetary Sciences, University of
  California, Riverside, CA 92521, USA}
\email{skane@ucr.edu}

\author[0000-0002-4860-7667]{Zhexing Li}
\affiliation{Department of Earth and Planetary Sciences, University of
  California, Riverside, CA 92521, USA}

%%%%%%%%%%%%%%%%%%%%%%%%%%%%%%%%%%%%%%%%%%%%%%%%%%%%%%%%%%%%%%%%%%%%

\begin{abstract}

Planetary rings are often speculated as being a relatively common
attribute of giant planets, partly based on their prevalence within
the Solar System. However, their formation and sustainability remain a
topic of open discussion, and the most massive planet within our
planetary system harbors a very modest ring system. Here, we present
the results of a N-body simulation that explores dynamical constraints
on the presence of substantial ring material for Jupiter. Our
simulations extend from within the rigid satellite Roche limit to 10\%
of the Jupiter Hill radius, and include outcomes from $10^6$ and
$10^7$ year integrations. The results show possible regions of a
sustained dense ring material presence around Jupiter that may
comprise the foundation for moon formation. The results largely
demonstrate the truncation of stable orbits imposed by the Galilean
satellites, and dynamical desiccation of dense ring material within
the range $\sim$3--29 Jupiter radii. We discuss the implications of
these results for exoplanets, and the complex relationship between the
simultaneous presence of rings and massive moon systems.

\end{abstract}


%%%%%%%%%%%%%%%%%%%%%%%%%%%%%%%%%%%%%%%%%%%%%%%%%%%%%%%%%%%%%%%%%%%%

\section{Introduction}
\label{intro}

A distinctive common feature of the giant planets within the Solar
System is the presence of ring systems orbiting the planet. Rings
systems have been detected and studied extensively for each of Jupiter
\citep{showalter1987,porco2003}, Saturn
\citep{pollack1975d,porco2005a}, Uranus \citep{elliot1977c,tyler1986},
and Neptune \citep{lane1989,showalter2020}. In particular, the
prominent rings of Saturn have been the source of numerous
investigations with regards to their formation
\citep{goldreich1978c,charnoz2009b} and dynamics
\citep{goldreich1978b,bridges1984}. For example, density waves
detected within Saturn's rings have been utilized as an effective
means to infer oscillations within the planetary interior
\citep{hedman2013c} and determine differential rotation of the outer
envelope \citep{mankovich2019} and a diffuse core
\citep{mankovich2021}. Additionally, the age of Saturn's rings has a
range of estimated values, from several hundred Myrs
\citep{zhang2017h,dubinski2019,iess2019} to as old as the Solar System
\citep{crida2019}. Age determinations depend on many factors, such as
the velocity dispersion and interaction between ring particles
\citep{salo1995}. Rings also stand as a possible record of past
collision events, and indeed the formation of Saturn's rings has been
suggested as the result of a substantial moon being desiccated by
tidal forces as it spiraled into the planet \citep{canup2010}.

By comparison, the Jupiter system contains a substantially more modest
ring system that has been extensively studied via data from such
missions as Voyager and Galileo, as well as ground-based observations
\citep{smith1979b,ockbertbell1999}. Theories regarding the origin and
evolution of the Jovian rings vary, such as their possible formation
along with the Galilean satellites \citep{prentice1979b}, and the
contributions of collisional ejecta lost from inclined satellites
\citep{burns1999} and escaping ejecta from the Galilean satellites
and/or the inner small moons
\citep{burns1999,esposito2002,krivov2002a}. Further possible sources
of potential ring material originate from impact debris
\citep{ahrens1994a,hueso2013,sankar2020} and the tidal disruption of
satellites \citep{hyodo2017b} or large passing Kuiper belt objects
\citep{hyodo2017a}. Such events are likely stochastic in nature whose
frequency is an age-dependent phenomenon
\citep{horner2008a}. Moreover, the dynamical evolution of Jupiter's
rings can have complicated explanations, partially with interactions
between ions and the Jovian magnetosphere \citep{horanyi1996b} and the
incorporation of dissipation effects \citep{greenberg1983a}. The
formation and dynamical evolution of Jupiter's rings have important
consequences regarding the prevalence of rings of giant
exoplanets. Exoplanetary rings can be a source of confusion when
evaluating the true nature of the planet and its properties
\citep{piro2018c,piro2020a} and their successful detection may reveal
important information regarding the formation of the planet and its
local environment \citep{arnold2004c,zuluaga2015a,sucerquia2020b}.
The Jupiter and Saturn systems demonstrate that the presence and
sustainability of rings may be an intricate function of the
architecture of planetary moons, as well as the intrinsic properties
of the planet itself. The discoveries and/or limits on exoplanet rings
and moons will provide crucial statistical data to further understand
how rings may have formed and evolved in our Solar System
\citep{kenworthy2015b}.

In this work, we provide the results of an extensive dynamical
simulation that tests regions of long-term dynamical stability for
ring systems near the plane of the Jovian equator, and in the presence
of the Galilean moons. In Section~\ref{arch}, we describe the
architectures of the Jupiter and Saturn systems, specifically the
structure of their rings with respect to the orbits of their
respective moons. Section~\ref{dynamics} provides the methodology and
results for our dynamical simulation of stable orbits at locations
that extend from within the rigid satellite Roche limit to 10\% of the
Jupiter Hill radius. The implications of our simulation results are
discussed in Section~\ref{discussion}, both in terms of the potential
for past/future Jovian ring systems and within the context of giant
exoplanets. Concluding remarks and suggestions for observational tests
are provided in Section~\ref{conclusions}.

%%%%%%%%%%%%%%%%%%%%%%%%%%%%%%%%%%%%%%%%%%%%%%%%%%%%%%%%%%%%%%%%%%%%

\section{Architecture of Jupiter and Saturn Systems}
\label{arch}

At the time of writing, the Solar System is known to contain over 200
moons, with Jupiter and Saturn harboring at least 79 and 82 moons,
respectively\footnote{\tt
  https://solarsystem.nasa.gov/moons/in-depth/}. The prevalence of
moons within the Solar System has been a primary motivator behind the
study of and search for exomoons
\citep[e.g.,][]{hinkel2013b,kipping2013d,heller2014c,hill2018}. It has
further been suggested that Solar System regular moons may serve as
analogs of compact exoplanetary systems in terms of their formation
and architectures
\citep{kane2013e,makarov2018,dobos2019,batygin2020b}. Regular moons
are particularly notable in that they likely formed with the planet,
as evidenced by their equatorial prograde orbits, and are often large
enough to exhibit hydrostatic equilibrium, resulting in a near-round
morphology. For example, the four Galilean moons likely formed from
the protoplanetary disk surrounding Jupiter
\citep{ogihara2012,heller2015e}, possibly catalyzed by migration of
Saturn \citep{ronnet2018}, and now contain $\sim$99.997\% of the total
mass orbiting the planet. The mass contained within the regular moons
will therefore have a significant influence on the dynamics of ring
formation and sustainability.

\begin{figure*}
  \begin{center}
    \includegraphics[angle=270,width=16.0cm]{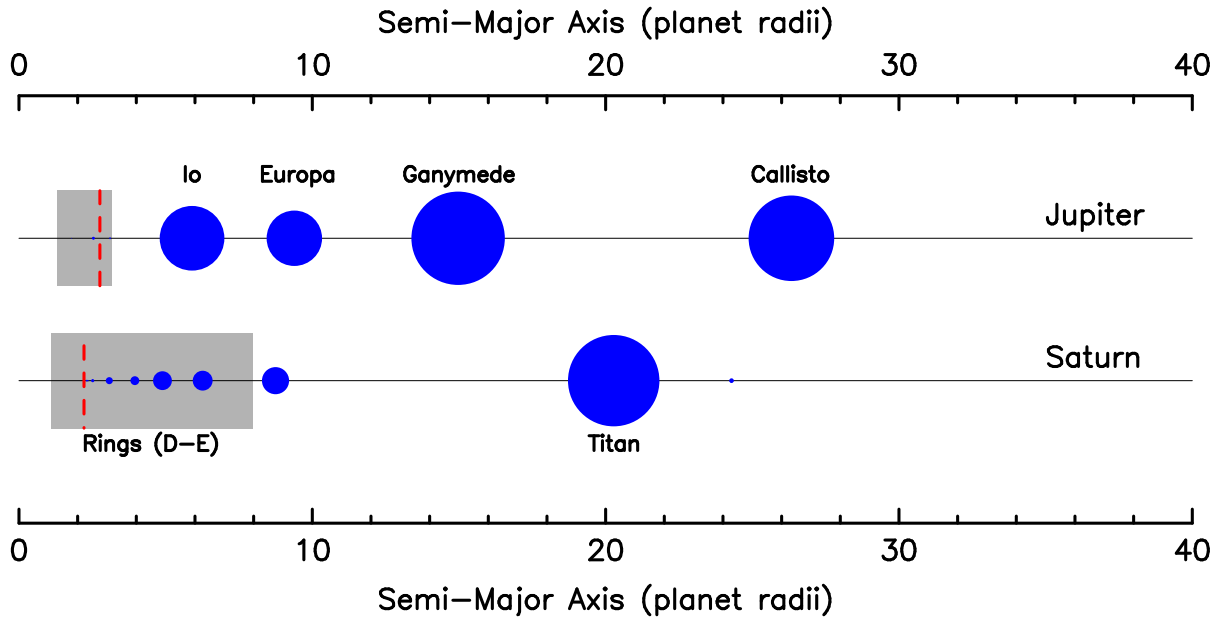}
  \end{center}
  \caption{The regular moons and rings of the Jupiter and Saturn
    systems, where the relative sizes of the moons are shown, and
    their semi-major axes are provided in units of the host planet
    radii. The extent of the current ring systems, including dense and
    tenuous rings, are shown as gray regions, and the vertical red
    dashed lines are the fluid satellite Roche limit for each planet.}
  \label{fig:radii}
\end{figure*}

Figure~\ref{fig:radii} provides a scaled view of the Jupiter and
Saturn systems, where the separation from the planetary centers,
located at zero, are provided in units of the respective host planet
radius. The regular moons are shown as blue circles, where their sizes
are relative to each other rather than in units of planetary radii,
and the names of the major moons are provided. It is worth noting that
all of Jupiter's regular moons, including the Galilean moons, have
semi-major axes that are within 1/25 of the Jupiter Hill radius,
possibly a result of the mass distribution of the circumplanetary disk
and moon migration processes that may have occurred. The extent of the
rings in units of the planetary radii are shown as gray regions, and
the vertical red dashed lines indicate the location of the fluid
satellite Roche limit for each planet. The depiction of Jupiter's
rings include the Halo to Thebe gossamer rings, which span a distance
of 1.29--3.16 Jupiter radii. Similarly, the depiction of Saturn's
rings include rings D--E, which extend a distance of 1.11--7.96 Saturn
radii. The fluid satellite Roche limits are 2.76 and 2.22 planet radii
for Jupiter and Saturn, respectively. Note that the rings depicted in
Figure~\ref{fig:radii} contain both ``dense'' and ``tenuous'' rings,
both of which have different sources of material and physical
processes acting upon them \citep{daisaka2001}. For example, Saturn's
E ring originates mainly from the cryovolcanic plumes of Enceladus,
and is considerably more tenuous than the main ring
\citep{horanyi2009,cuzzi2010a}.

Figure~\ref{fig:radii} highlights the differences in the distribution
of moon mass with respect to the rings for each of the planets. In
particular, the rings of Saturn contain numerous small moons, whose
presence can both feed the rings with new material, and also
``shepherd'' the rings through their gravitational influence
\citep{petit1988c,charnoz2011b,cuzzi2014b,nakajima2020}. Indeed, moons
can accrete relatively rapidly from ring material beyond the Roche
limit, and such rings are thought to have been a source of numerous
moons within the present Saturn architecture
\citep{charnoz2010,crida2012,salmon2017}. Gaps in Saturn's rings form
through several processes, among them orbital resonances with moons,
such as the relationship between the Cassini Division and Mimas
\citep{goldreich1978c,iess2019,noyelles2019}. Titan is relatively far
from the main ring structure, but its presence does result in a
ringlet within the inner C ring through the effect of orbital
resonance \citep{porco1984a}. However, there is evidence to suggest
that Titan has experienced an outward migration through tidal
dissipation \citep{lainey2020}. This is somewhat in contrast to the
effects of the more massive Galilean moons on the dynamical
environment around Jupiter, particularly as Io, Europa, and Ganymede
are located in a 4:2:1 Laplace resonance
\citep{malhotra1991,peale2002b}. Their orbital configuration is
interpreted as strong evidence that the moons migrated inward either
during formation or soon thereafter
\citep{greenberg1987,peale2002b,sasaki2010,ogihara2012}. This may have
resulted in a significant gravitational truncation of a potential
massive ring system for Jupiter, such as that seen for Saturn,
depending on the formation and migration timescales of the Galilean
moons. Overall, there are significant differences between the ways in
which dense and tenuous rings interact with the planet and satellite
system, partially depending on the location of the rings relative to
the Roche limits. The simulations described in this paper consider
only dense rings for a range of semi-major axes. The physical
processes acting upon the rings are described further in
Section~\ref{rings}.

%%%%%%%%%%%%%%%%%%%%%%%%%%%%%%%%%%%%%%%%%%%%%%%%%%%%%%%%%%%%%%%%%%%%

\section{Dynamical Simulations}
\label{dynamics}

This section describes the dynamical simulations, including their
configuration, results for the Galilean moons, and injection of ring
particles.

%%%%%%%%%%%%%%%%%%%%%%%%%%%%%%%%%%%%%%%%%%%%%%%%%%%%%%%%%%%%%%%%%%%%

\subsection{Dynamics of the Galilean Moons}
\label{galilean}

As discussed in Section~\ref{arch}, an important factor in the
formation and evolution of planetary rings is the moons present in the
system. In this context, the Galilean moons are by far the largest
gravitational influence on the presence of rings within the inner part
of the Jupiter system. Thus, the dynamical evolution of the Galilean
moons is a crucial component of evaluating potential ring
sustainability. There are substantial data that contribute toward
accurate ephemerides of the moons, and their dynamics have previously
been studied in detail
\citep{lieske1980,greenberg1987,lainey2004a,lari2018}. However, the
vast majority of these studies focus on the short-term
($\sim$100~year) dynamics of the moons, whereas this study is
concentrated on timescales related to the sustainability of planetary
rings (1--10 million years).

\begin{figure*}
  \begin{center}
    \includegraphics[angle=270,width=16.0cm]{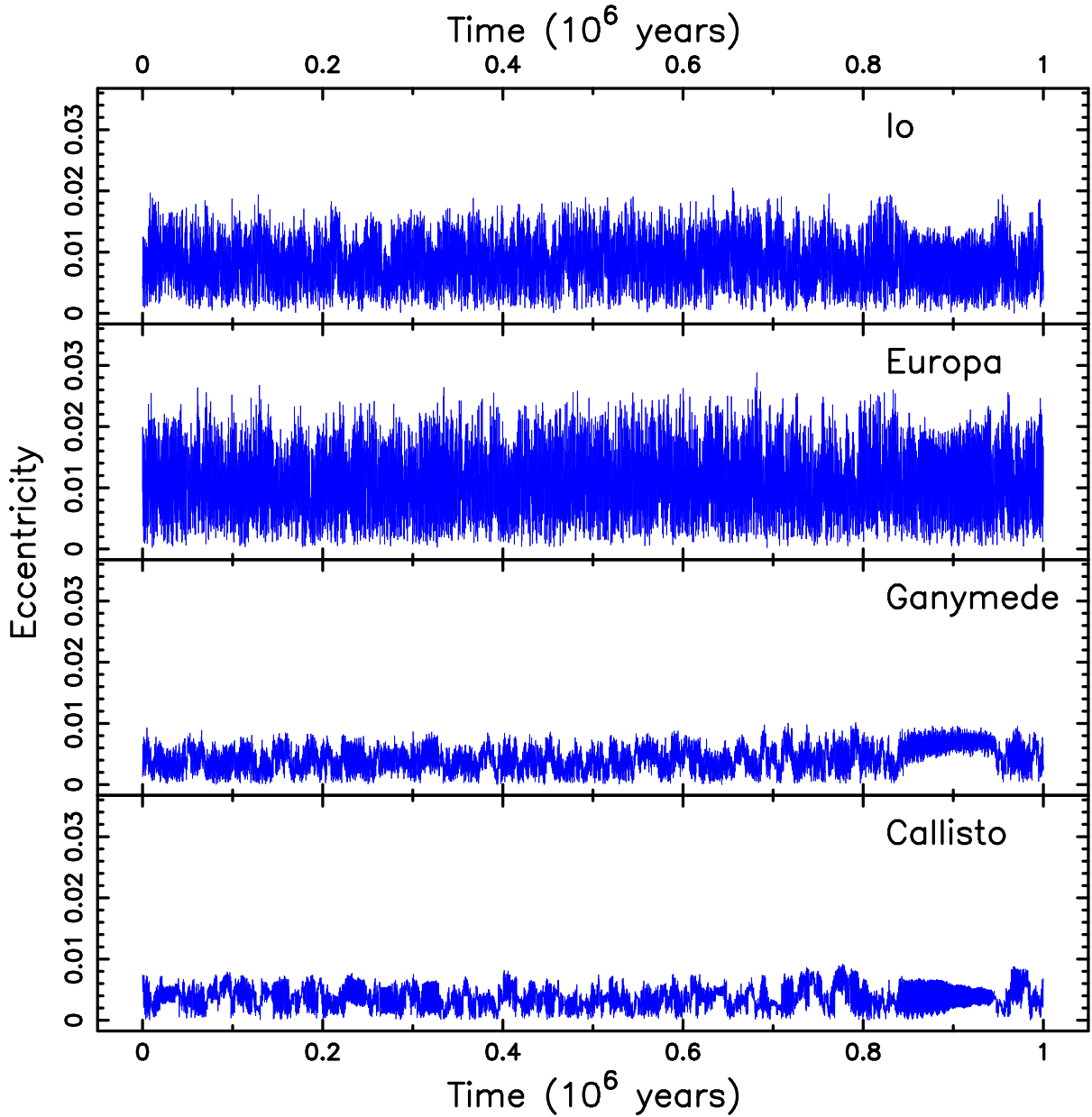}
  \end{center}
  \caption{Eccentricity as a function of time for the Galilean
    satellites: Io, Europa, Ganymede, and Callisto. The eccentricity
    variations were recorded every 100 years during a $10^6$~year
    simulation, as described in Section~\ref{dynamics}.}
  \label{fig:ecc}
\end{figure*}

The simulations carried out for this work were conducted within the
dynamical simulation package REBOUND \citep{rein2012a} with the
symplectic integrator WHFast \citep{rein2015c}. The initial conditions
of the system were constructed to reproduce the configuration of the
Galilean moon system, incorporating the current orbital elements
extracted from the Horizons DE431 ephemerides \citep{folkner2014}, and
including the effects of Jupiter's oblateness \citep{tamayo2020a}. The
dynamical simulation presented in this work provides an independent
assessment for the eccentricity evolution of the Galilean moons, the
results of which are shown in Figure~\ref{fig:ecc} for a duration of
$10^6$~years. The combination of the resonance configuration and tidal
dissipation present in the Galilean system ensures that the moons have
exceptionally stable orbits through time and that they remain largely
circular. Note that tidal dissipation was not included in our
dynamical simulation. Figure~\ref{fig:ecc} shows that the
eccentricities remain below 2\%, 3\%, 1\%, and 1\% for Io, Europa,
Ganymede, and Callisto, respectively. The short-term analyses of the
Galilean moon eccentricities, such as the semi-analytical model
provided by \citet{lari2018}, reveal a transfer of angular momentum
within the Laplace resonance of the three inner moons with a period in
the range 400-500 days. The long-term eccentricity evolution
represented in Figure~\ref{fig:ecc} does not reveal periodic behavior
at significantly longer timescales, but does demonstrate that angular
momentum transfer results in higher eccentricity variations for the
less massive inner two moons than the more massive outer two
moons. The slightly higher eccentricity variations for Io and Europa
increases regions of dynamical instability in their vicinity, and thus
has consequences for long-term ring stability close to Jupiter.

%%%%%%%%%%%%%%%%%%%%%%%%%%%%%%%%%%%%%%%%%%%%%%%%%%%%%%%%%%%%%%%%%%%%

\subsection{Particle Injection and Ring Stability}
\label{particle}

To explore the gravitational constraints that the Galilean moons
impose upon potential ring particles orbiting Jupiter, we conducted a
suite of dynamical simulations for the system based upon the
architecture framework described in Section~\ref{galilean}. The
stability of an extensive ring system and possible moon forming
material around Jupiter was tested by introducing a series of test
particles to the Galilean moon system and evaluating their
survivability at different locations within the system. The test
particles were assumed to have a density of water ice
(0.917~g/cm$^3$), the value for which is representative of most of the
materials in Saturn's ring system, and a spherical shape with radius
of $\sim$1~meter, yielding a total particle mass of
$\sim$3841~kg. Thus, we are considering only dense ring material and
the gravitational perturbations acting upon them (see
Section~\ref{rings}). The particle orbits were set to be circular with
an inclination of $\sim$0\degr\ with respect to the Jupiter equator
(coplanar with the Galilean moons). The test particles were placed at
different locations from Jupiter, extending from inside the rigid
satellite Roche limit (see Figure~\ref{fig:radii}) to $1/10$ of the
Jupiter Hill radius. This region was sampled with 1000 evenly spaced
locations, resulting in a location step size of $\sim$4963~km. The
simulations were carried out with both $10^6$~year and $10^7$~year
integration times for each of the separation cases, which translates
to $4.6 \times 10^6$ and $4.6 \times 10^7$ orbits at the outer edge of
our sample space, respectively. For each simulation, the time step was
set to be 0.05 of Io's orbital period (2.1~hours), except for cases
where the test particle locations were inside Io's orbit, where the
time step was adjusted to 0.05 of the orbital period at each particle
location.  This ensured an adequate time resolution to properly sample
the dynamical interactions between test particles and the Galilean
moons. The orbital properties of the particles and Galilean moons were
output every 100 years during the simulation and the survival rates of
the test particle were calculated at each orbital separation. Test
particle survival was based upon the elapsed time to be either ejected
from the system or suffering a collision with one of the large system
bodies during the entire simulation.

\begin{figure*}
  \begin{center}
    \includegraphics[angle=270,width=16.0cm]{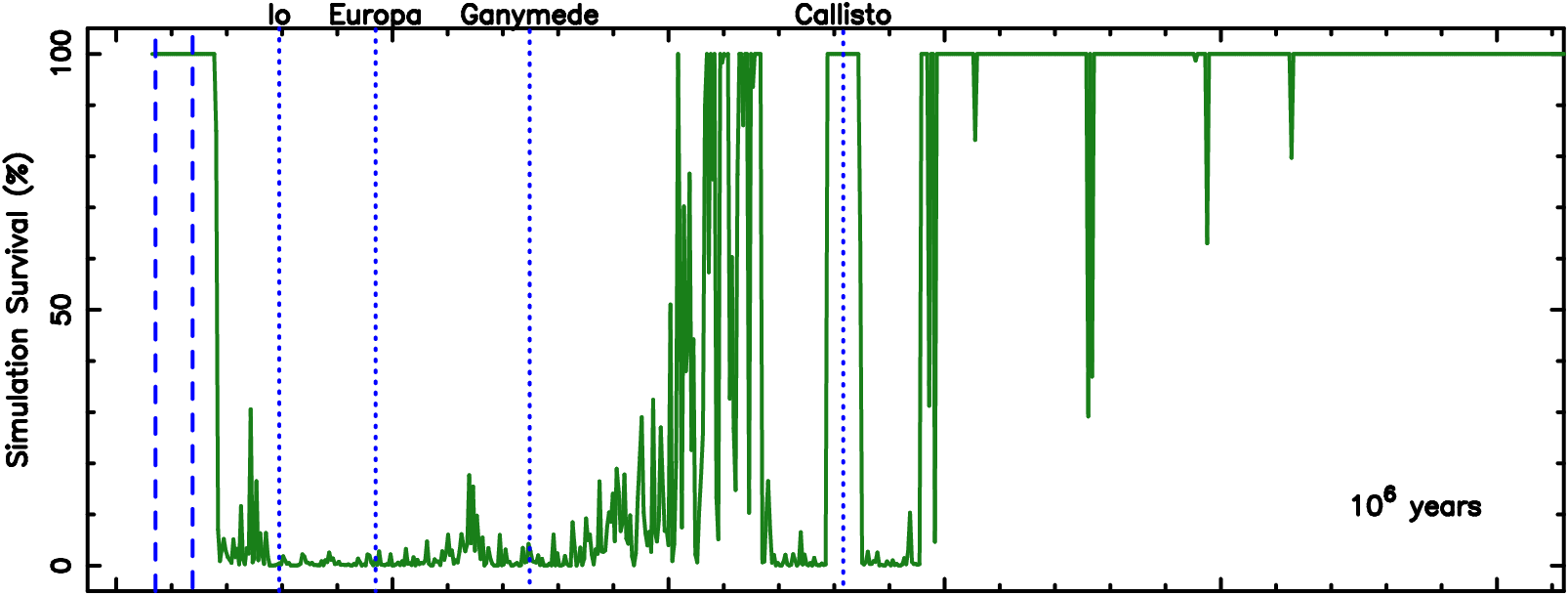} \\
    \includegraphics[angle=270,width=16.0cm]{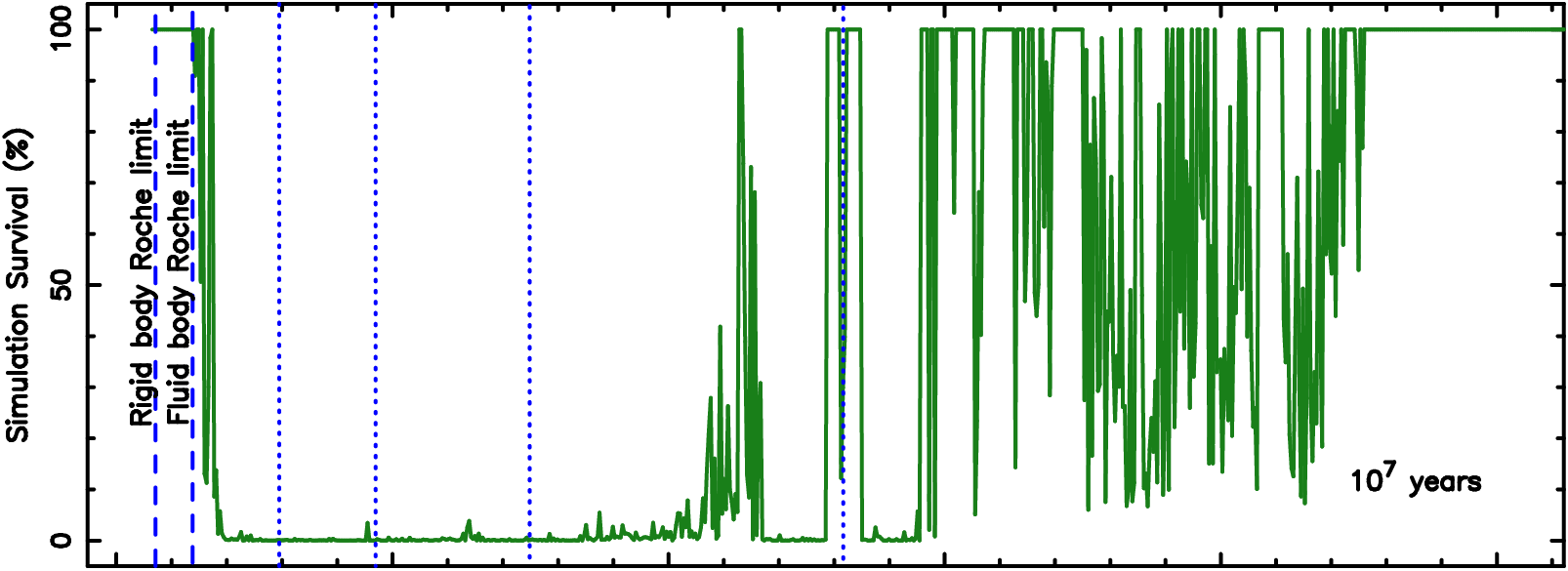} \\
    \includegraphics[angle=270,width=16.0cm]{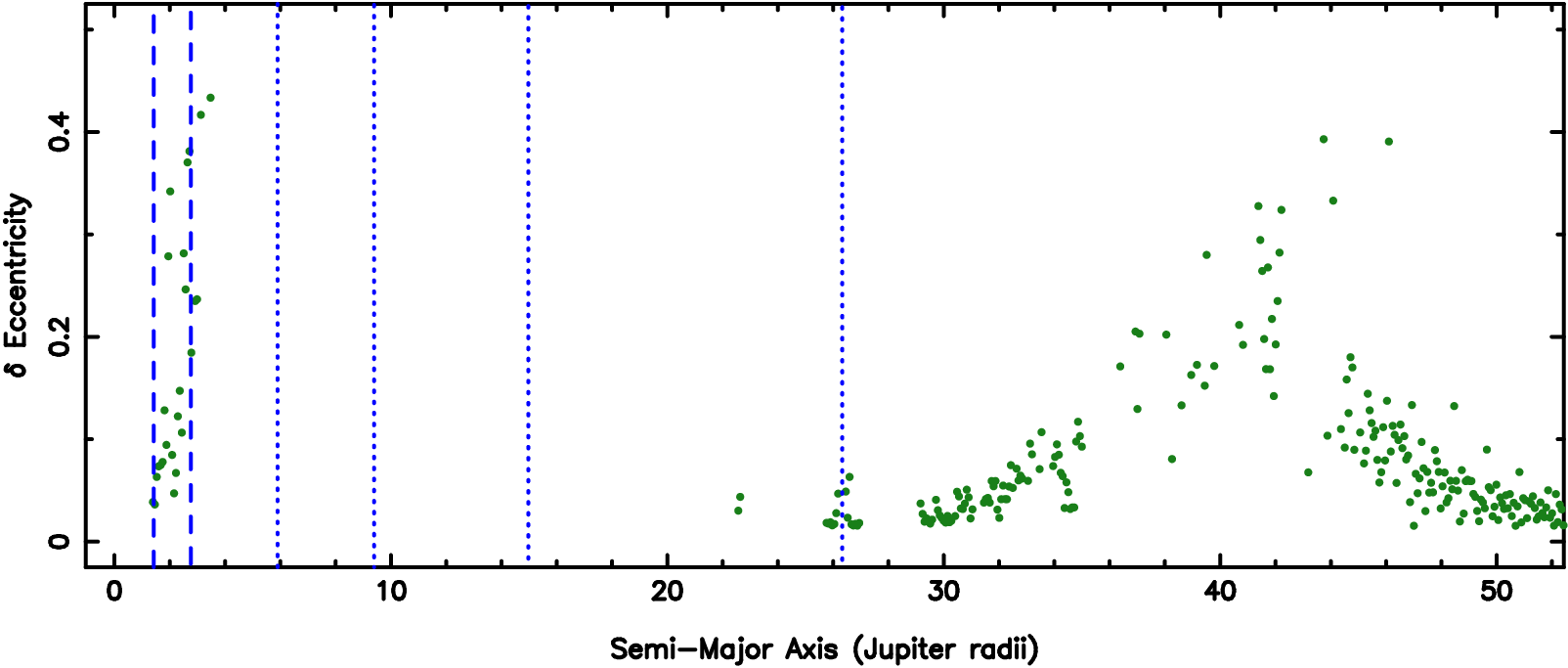}
    \end{center}
  \caption{Results of particle injection and survival for the $10^6$
    year (top panel) and $10^7$ year (middle panel) dynamical
    simulations. The horizontal axis is the separation from Jupiter in
    planetary radii, and the vertical axis shows the percentage of the
    total dynamical simulation that particles survived at that
    location, represented by the green line. The vertical dotted lines
    represent the semi-major axes of the Galilean moons, and the
    vertical dashed lines represent the locations of the rigid and
    fluid Roche limits. The bottom panel shows the change in
    eccentricity that occurs for each particle as a function of their
    initial semi-major axis during the course of the $10^7$ year
    simulations.}
  \label{fig:sim}
\end{figure*}

The primary results of our simulations that include particle
injections are shown in the top two panels of Figure~\ref{fig:sim},
and are represented as the percentage of the total simulation survived
by particles for each separation location. The simulation durations of
$10^6$~years and $10^7$ years are shown in the top and bottom panels,
respectively, and the location of the Galilean moons are indicated by
vertical dotted lines. The vertical dashed lines indicate the location
of both the rigid and fluid body Roche limits. The top panel shows
that, even for $10^6$ years, the region surrounding the Laplace
resonance of Io, Europa, and Ganymede is rendered unstable,
exacerbated by the slight eccentricity variations of Io and Europa
described in Section~\ref{galilean}. The inner limits of dynamical
stability are located at $\sim$3.55 Jupiter radii and $\sim$3.12
Jupiter radii for the $10^6$~year and $10^7$~year simulations,
respectively. The current Jupiter ring system, depicted in
Figure~\ref{fig:radii}, extends to $\sim$3.16 Jupiter radii,
indicating that the ring is young ($< 10^7$~years) and/or is being
sustained with additional material to the Thebe extension
\citep{borisov2021c}. The structure of the dynamical stability beyond
the orbit of Callisto is a complicated result of Ganymede and Callisto
resonance locations. For example, the dynamical instability located at
$\sim$31.2 Jupiter radii, partially present at $10^6$~years and
significantly more pronounced at $10^7$~years, is the 1:3 resonance
location with Ganymede. Furthermore, there is an island of stability
located at $\sim$41.8 Jupiter radii, which is not evident in the
$10^6$~year results but becomes apparent at $10^7$~years, which is a
result of a 1:2 resonance with Callisto. The effect of resonances is
further emphasized in the bottom panel of Figure~\ref{fig:sim}, which
shows the change in eccentricity, $\delta e$, that occurs during the
course of the $10^7$ year simulations. It is worth noting that,
although the bottom panel of Figure~\ref{fig:sim} implies that
particles close to the Roche limits are predominantly scattered into
high eccentricity orbits, stable low-eccentricity orbits can exist
within that region. For example, the inner Jovian moons of Metis,
Adrastea, Amalthea, and Thebe have semmi-major axes that lie within
the range 1.8--3.2 Jupiter radii and are known to have eccentricities
that are below 0.02 \citep{cooper2006,borisov2020}, although dynamical
interactions with Io have inflated the eccentricity of Thebe
\citep{burns2004a}. We conducted additional simulations for the
specific cases of those four inner moons where, as for the Galilean
moons, current orbital elements were extracted from the Horizons DE431
ephemerides. The results of these simulations show that their mean
eccentricities remained small throughout the entire simulation, with
mean eccentricities of 0.008, 0.008, 0.005, and 0.016 for Metis,
Adrastea, Amalthea, and Thebe, respectively. As noted in
Section~\ref{galilean}, tidal dissipation was not included in our
simulations, and so these dynamically induced eccentricities may be
considered upper limits to their expected values.

\begin{figure}
  \includegraphics[angle=270,width=8.5cm]{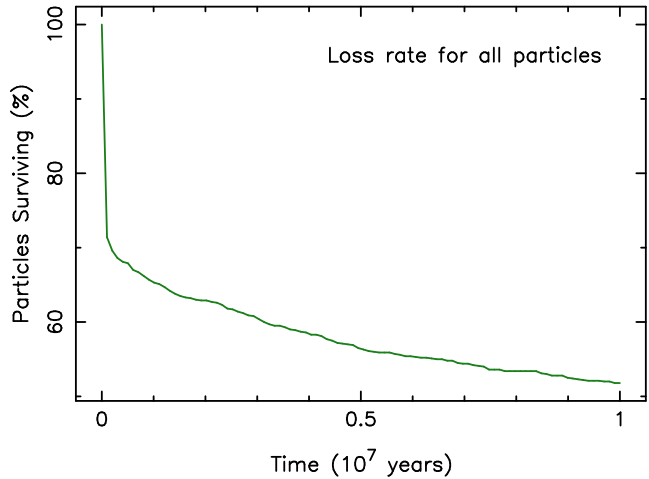}
  \caption{The loss rate for all particles within the investigated
    semi-major axis region around Jupiter, shown as a function of time
    (fractions of $10^7$~years) and the percentage of particles that
    survive to that time.}
  \label{fig:lossrate}
\end{figure}

The rate at which particles are
lost from within the investigated semi-major region is shown in
Figure~\ref{fig:lossrate}. Almost 30\% of the particles are lost
within the first $10^5$~years of simulation time. After $10^6$~years,
$\sim$65\% of particles remain and, after $10^7$~years, $\sim$52\% of
particles remain. As noted above, lost particles are generally
considered to be the result of ejection or collisions. However, as
described in Section~\ref{arch}, particles beyond the fluid body Roche
limit may also coalesce into moons on relatively short timescales, and
the diversity of eccentricities exhibited in the bottom panel of
Figure~\ref{fig:sim} for many of the surviving particles supports this
possible outcome. Another source of particle loss can be capture by
the Galilean moons, forming circumsatellital rings
\citep{sucerquia2021}. Overall, the Galilean moons carve a substantial
area of instability into the region around Jupiter that may only allow
relatively short-lived ring systems to co-exist with the orbital
architectures.

%%%%%%%%%%%%%%%%%%%%%%%%%%%%%%%%%%%%%%%%%%%%%%%%%%%%%%%%%%%%%%%%%%%%

\section{Discussion}
\label{discussion}

%%%%%%%%%%%%%%%%%%%%%%%%%%%%%%%%%%%%%%%%%%%%%%%%%%%%%%%%%%%%%%%%%%%%

\subsection{Implications for Jovian Rings}
\label{rings}

As the dominant planetary mass in the Solar System, Jupiter has
undoubtedly had a complex and eventful dynamical history. Such a
history includes significant impacts and other events that have
resulted in raw material for substantial ring formation. The major
question regarding these ring structures pertain to their long-term
viability within the architecture of the system. The results of our
simulations described in Section~\ref{particle} demonstrate that ring
structures within the separation range $\sim$3--29 Jupiter radii are
mostly removed within $10^6$~years. What remains are a system of
ringlets between Ganymede and Callisto ($\sim$20--24 Jupiter radii),
and rings that coincide with the orbit of Callisto. Rings beyond 29
Jupiter radii are viable for $10^6$~years, but such rings would likely
consist of remaining material from planet and moon formation processes
when the Solar System was relatively young. After $10^7$~years, the
dynamical influence of the Galilean moons further desiccates the
material in the region of $\sim$3--29 Jupiter radii, including the
ringlet structure between Ganymede and Callisto. As discussed in
Section~\ref{particle}, orbital resonances with Ganymede and Callisto
also compromise the region beyond 29 Jupiter radii, ensuring that the
primordial material would not have remained beyond a few tens of
millions of years.

The results presented in this work are based primarily on dynamical
interactions within the Jupiter system. However, there are numerous
other factors that have not been taken into account. Events that may
increase the lifetime of ring structures, such as impacts and
outgassing of moons that release solid particles, have not been
considered. Jupiter is well known to suffer relatively frequent
impacts \citep{zahnle2003a,horner2008a,hueso2013,hueso2018b}, such as
the Comet Shoemaker-Levy-9 event \citep{zahnle1994,asphaug1996b}, with
observations of recent impact events helping to place constraints on
ejection material that achieves escape velocity
\citep{ahrens1994a,sankar2020}, and thus may contribute to ring
material. It is worth noting that contributions from isotropic
impactors (compared with ecliptic impactors) may result in a
stochastic distribution of ring material, whereas our simulations
specifically investigate rings coplanar with Jupiter's
equator. Furthermore, orbital precession and collisions will naturally
cause any orbiting debris to collapse into the planet's equatorial
plane. Additional sources of ring material are the grinding down of
small moons and outgassing from the Galilean moons
\citep{burns1999,esposito2002}.

On the other hand, there are numerous processes that serve to decrease
the ring lifetime, such as Poynting-Robertson drag, the Yarkovsky
effect, and the electromagnetic influence of the Jovian magnetosphere
\citep{burns1999,rubincam2006,kobayashi2009a}. Such non-gravitational
forces primarily govern particles that are significantly smaller
($\mu$m--mm) than those used in our simulations \citep{burns1979c},
and so are unlikely to substantially impact the results presented
here. For dense rings, such as those that are considered in our
simulations, a major source of ring material loss beyond the Roche
limit is the accretion into small moons \citep{crida2012}. Such
accretion can happen on relatively small timescales, such that the
loss of ring material in this manner may occur well within the periods
of time considered by our simulations. In fact, our results provided
in Section~\ref{particle} demonstrate that the excitation of particles
into eccentric orbits may further promote collision
scenarios. However, moonlets that form in the investigated region of
semi-major axes will likely remain subject to the instability caused
by the Galilean moons, resulting in possible ejection from the
system. Migration of moons have also played a role in shaping the
dynamical environment around Jupiter. Resonances between moons can
promote inward and outward migration and is thought to have occurred
for both the Jupiter and Saturn moon systems \citep{fuller2016}. The
mechanics of resonance trapping for moons is the same process that
takes place for planetary migration \citep{wyatt2003e} and can greatly
affect their composition, such as the case of the TRAPPIST-1 planets
\citep{unterborn2018a}. Indeed, interaction with the rings themselves
can result in the migration of small moons \citep{bromley2013}, which
may in turn result in their contribution to the rings near the Roche
limit. Furthermore, given the relatively small mass of the test
particles used in our simulations, the calculated life expectancy of
those particles may be considered an upper limit in many cases.

With the various above described competing factors in ring
contributions, the sustainability of planetary rings may be sensitive
to the architecture of the local environment. Based on our
simulations, it is possible that the Galilean moons may be a
significant reason that, integrated over time, Jupiter is unable to
harbor substantial rings. However, there may also have been periods of
Jupiter history during which impact and small moon collision rates
exceeded dynamical disruption by the Galilean moons, allowing
sustained rings significantly more massive than those observed at the
present epoch. Although the sustainability of an extended ring system
beyond the Roche limit may be dominated by the combination of
dynamical effects and moon coalescence, the lifetime of rings within
the Roche limit is substantially more complicated. For example, in the
case of Saturn, an examination of the Pallene dusty ring found that
non-gravitational forces dominate over the dynamical effects with
regards to the sustainability of the ring, although the moon mass
within the Saturnian system is significantly less than that of the
Galilean moons \citep{munozgutierrez2022}. Therefore, the relative
weighting of ring sustainability and dessication factors close to the
planet have yet to be determined, and it is unclear that the dynamical
effects are the primary factor in that region.

%%%%%%%%%%%%%%%%%%%%%%%%%%%%%%%%%%%%%%%%%%%%%%%%%%%%%%%%%%%%%%%%%%%%

\subsection{Implications for Exoplanets}
\label{exoplanets}

\begin{figure}
  \includegraphics[angle=270,width=8.5cm]{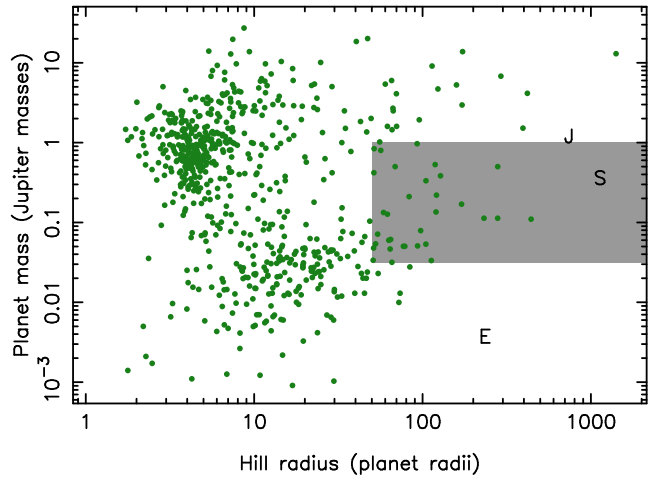}
  \caption{Planet masses (Jupiter masses) and Hill radii (planetary
    radii) for the known exoplanets with measured masses and
    radii. The locations of Earth, Jupiter, and Saturn are marked by
    E, J, and S, respectively. The gray region indicates where
    formation of massive moons may be limited by disk migration,
    whilst a larger Hill radius maximizes primordial and new material
    to contribute to ring formation.}
  \label{fig:exo}
\end{figure}

As noted in Section~\ref{intro}, rings are an important component of
planetary evolution, and there are active efforts to search for their
presence around exoplanets
\citep{arnold2004c,zuluaga2015a,sucerquia2020b}. It is expected that
primordial ring structures, inherited from the circumplanetary disk,
contain a mass that scales with the feeding zone, and hence the Hill
radius, of the planet \citep{heller2014c,kaib2015a}. There is also
evidence that moons play a significant role in this early ring
evolution, and that the disks are truncated well inside the extent of
the Hill radius \citep{shabram2013,fujii2017a}. At the present epoch,
the respective prevalence of rings around Jupiter and Saturn suggest
that post-primordial ring formation and evolution does not scale with
planet mass, but rather with the particular moon architecture
associated with the host planet. Indeed, the circumplanetary mass
required to form a Galilean moon system analog may result in a natural
inhibitor to ring formation around massive giant planets. In
particular, moon migration into a Laplace resonance, such as that
described in Section~\ref{arch}, produces strong destabilizing
perturbations that quickly desiccate disk particles within several
tens of planetary radii of the planet. It is further suggested by
\citet{canup2006} that disk migration may limit the maximum size of
moons, beyond which the moons may migrate through the circumplanetary
disk and into the planet. The implication is that Saturn may possibly
represent a ``sweet spot'' of ring formation in terms of mass, Hill
radius, and moon-forming capacity, although this is quite difficult to
gauge fully without more complete knowledge of the Jupiter and Saturn
architecture evolutions. Moreover, careful consideration must be given
to the various processes acting upon ring formation and dessication,
described in Section~\ref{rings}. In particular, there are numerous
non-gravitational forces acting upon ring material within or close to
the Roche limit that can have a dominating effect on ring evolution
(see Section~\ref{rings}). It is also worth noting that these implied
correlations between planet mass and ring/moon prevalence are based
upon the limited inventory of giant planets present in the Solar
System, for which exoplanet studies will provide a much needed
statistical validation.

However, if indeed the presence of relatively massive moon systems is
correlated with planet mass and Hill radius, then the prevalence of
ring systems may be likewise rare amongst many of the discovered giant
exoplanets. Shown in Figure~\ref{fig:exo} are planetary masses
(Jupiter masses) and calculated Hill radii (planetary radii) for known
exoplanets with relavant data available. The data were extracted from
the NASA Exoplanet Archive \citep{akeson2013} and are current as of
2021 December 18 \citep{nea}. For comparison, the locations of Earth,
Jupiter, and Saturn are also marked on the diagram. The gray box
provides a rough guide for what may be considered an optimization
region where the harboring of substantial planetary rings could be
favorable. The planet mass range of the gray box is 10 Earth masses to
1 Jupiter mass to approximately capture planets within the ice/gas
giant regime
\citep{weiss2013,lammer2014a,lopez2014,rogers2015a,chen2017} whilst
limiting the potential for substantial moon formation. The Hill radius
range is for all Hill radii beyond 50 planetary radii since small Hill
radii truncates both the extent of the circumplanetary disk
\citep{shabram2013} and the dynamical viability of moons
\citep{barnes2002b,hinkel2013b,kane2017c}. Saturn is deeply embedded
within the gray region, whilst Uranus and Neptune (also fulfilling the
planet mass criteria) are located at Hill radii of 2613 and 4644
planetary radii, respectively. As described above, planets more
massive than Jupiter, with larger mass proto-satellite disks and a
larger Hill radius in which to engage in moon formation, may
experience significantly reduced timescale of ring
sustainability. Such a timescale reduction would result from the rapid
formation of large moons, whose gravitational presence would, in turn,
either eject the remaining ring material or excite their
eccentricities resulting in further enhancement of moon formation (see
Section~\ref{particle}). The size of the box shown in
Figure~\ref{fig:exo} is empirical in nature and requires further
investigation of the various competing factors toward ring formation
and sustainability, but may serve as a useful guide for testing models
regarding how the presence of substantial moons could potentially
influence the long-term presence of rings around giant planets.

%%%%%%%%%%%%%%%%%%%%%%%%%%%%%%%%%%%%%%%%%%%%%%%%%%%%%%%%%%%%%%%%%%%%

\section{Conclusions}
\label{conclusions}

Planetary rings and moons are very important features of our Solar
System, both in their intrinsic geologic and dynamical properties, and
as crucial signposts of planetary formation and
evolution. Understanding the complex interactions between moons and
rings, and how these vary with planetary mass, composition, Hill
radius, and time, remains one of the most intricate research topics in
planetary science. The dynamical evolution of giant planet systems is
one of the primary ways in which tracing of rings systems and their
ages may be undertaken.

The results of our dynamical simulations demonstrate that the presence
of massive moons, especially systems that have migrated into resonance
traps as for the Galilean moons, can create significant dynamical
constraints on ring systems comprised of dense material. This
indicates that, although Jupiter may have had intermittent periods of
substantial rings systems, their long-term sustainability may be
severely truncated by the presence of the Galilean moons and
associated resonances. Ring material beyond the Roche limit that
remain in stable orbits may also experience eccentricity excitation
that enhances moonlet coalescence. Furthermore, we have shown that the
outer edge of the present ring system must be relatively young ($<
10^7$~years) in order to have survived dynamical scattering
processes. The balance between planet mass, the formation of massive
moons, and the sustainability of significant ring mass, means that
Saturn may be near the optimal region for the formation and long-term
survival of substantial rings. A useful extension to this work could
thus include longer timescale simulations combined with migration
effects that fully explore the interaction between moons and rings
during periods of formation, as well as moon formation from ring
material and the inclusion of non-gravitational forces near the Roche
limit.

Although the inventory of Solar System giant planets and their
associated rings and moons is limited, they yet provide the best clues
to the formation and evolution of such systems
\citep{horner2020b,kane2021d}, as well as a guide toward detecting
their exoplanet analogs
\citep{dalba2015,mayorga2016,mayorga2020,wakeford2020b}. The detailed
data available for local giant planet systems must necessarily be
balanced by the statistical knowledge that will be gained through the
discovery of exomoons and exorings. Such discoveries will provide the
means to fully explore the above described potential correlation of
moons and ring properties with those of their host planet.

%%%%%%%%%%%%%%%%%%%%%%%%%%%%%%%%%%%%%%%%%%%%%%%%%%%%%%%%%%%%%%%%%%%%

\section*{Acknowledgements}

The authors thank Paul Dalba and the anonymous referees for useful
feedback on the manuscript. This research has made use of the NASA
Exoplanet Archive, which is operated by the California Institute of
Technology, under contract with the National Aeronautics and Space
Administration under the Exoplanet Exploration Program. The results
reported herein benefited from collaborations and/or information
exchange within NASA's Nexus for Exoplanet System Science (NExSS)
research coordination network sponsored by NASA's Science Mission
Directorate.

%%%%%%%%%%%%%%%%%%%%%%%%%%%%%%%%%%%%%%%%%%%%%%%%%%%%%%%%%%%%%%%%%%%%

\software{REBOUND \citep{rein2012a}}

%%%%%%%%%%%%%%%%%%%%%%%%%%%%%%%%%%%%%%%%%%%%%%%%%%%%%%%%%%%%%%%%%%%%

%\bibliographystyle{aasjournal}
%\bibliography{/data/skane/latex/styles/references,nea}

%%%%%%%%%%%%%%%%%%%%%%%%%%%%%%%%%%%%%%%%%%%%%%%%%%%%%%%%%%%%%%%%%%%%

\end{document}